\documentclass[prl,letterpaper,onecolumn,showpacs,superscriptaddress,amsmath,amsfonts,amssymb,floatfix,preprintnumbers,citeautoscript]
{revtex4}
\pdfoutput=1
\usepackage{dcolumn,graphicx,color,booktabs} \graphicspath{{Images/}}
\usepackage{amsmath,bm}
\usepackage[outercaption]{sidecap}
\renewcommand{\figurename}{Fig.}
\makeatletter\renewcommand{\fnum@figure}[1]{\figurename~\thefigure.}\makeatother
\definecolor{DarkBlue}{rgb}{0,0,0.5}

\begin{document} \pagestyle{plain}


\title{Momentum-resolved superconducting gap in the bulk of Ba$_{1-x}$K$_{x}$Fe$_2$As$_2$ from combined ARPES and $\mu$SR measurements}

\author{D.\,V.~Evtushinsky}
\affiliation{Institute for Solid State Research, IFW Dresden, P.\,O.\,Box 270116, D-01171 Dresden, Germany}
\author{D.\,S.~Inosov}
\affiliation{Institute for Solid State Research, IFW Dresden, P.\,O.\,Box 270116, D-01171 Dresden, Germany}
\affiliation{Max-Planck-Institute for Solid State Research, Heisenbergstrasse 1, D-70569 Stuttgart, Germany}
\author{V.~B.~Zabolotnyy}
\affiliation{Institute for Solid State Research, IFW Dresden, P.\,O.\,Box 270116, D-01171 Dresden, Germany}
\author{M.\,S.\,Viazovska}
\affiliation{Max-Planck-Institute for Mathematics, Vivatsgasse 7, 53111 Bonn, Germany}
\author{R.\,Khasanov}
\affiliation{Laboratory for Muon Spin Spectroscopy, Paul Scherrer Institut, CH-5232 Villigen PSI, Switzerland}
\author{A.~Amato}
 \affiliation{Laboratory for Muon Spin Spectroscopy, Paul Scherrer
Institut, CH-5232 Villigen PSI, Switzerland}
\author{H.-H.~Klauss}
 \affiliation{IFP, TU Dresden, D-01069 Dresden, Germany}
\author{H.~Luetkens}
 \affiliation{Laboratory for Muon Spin Spectroscopy, Paul Scherrer
Institut, CH-5232 Villigen PSI, Switzerland}
\author{Ch.~Niedermayer}
 \affiliation{Laboratory for Neutron Scattering, Paul Scherrer Institute and ETH
 Z\"urich,CH-5232 Villigen PSI, Switzerland}
\author{G.\,L.\,Sun} \author{V.~Hinkov}
\affiliation{Max-Planck-Institute for Solid State Research, Heisenbergstrasse 1, D-70569 Stuttgart, Germany}
 \author{C.\,T.~Lin}
 \affiliation{Max-Planck-Institute for Solid State Research, Heisenbergstrasse 1, D-70569 Stuttgart, Germany}
\author{A.~Varykhalov}
\affiliation{BESSY GmbH, Albert-Einstein-Strasse 15, 12489 Berlin, Germany}
\author{A.\,Koitzsch}\author{M.~Knupfer}\author{B.~Büchner}
\affiliation{Institute for Solid State Research, IFW Dresden, P.\,O.\,Box 270116, D-01171 Dresden, Germany}
\author{A.\,A.~Kordyuk}
\affiliation{Institute for Solid State Research, IFW Dresden, P.\,O.\,Box 270116, D-01171 Dresden, Germany}
\affiliation{Institute of Metal Physics of National Academy of Sciences of Ukraine, 03142 Kyiv, Ukraine}
\author{S.\,V.~Borisenko}
\affiliation{Institute for Solid State Research, IFW Dresden, P.\,O.\,Box 270116, D-01171 Dresden, Germany}

\begin{abstract}
\noindent Here we present a calculation of the temperature-dependent London penetration depth, $\lambda(T)$, in Ba$_{1-x}$K$_{x}$Fe$_2$As$_2$ (BKFA) on the basis of the electronic band structure \cite{Volodya, Volodya2} and momentum-dependent superconducting gap \cite{Evtushinsky} extracted from angle-resolved photoemission spectroscopy (ARPES) data. The results are compared to the direct measurements of $\lambda(T)$ by muon spin rotation ($\mu$SR) \cite{Khasanov}. The value of $\lambda(T=0)$, calculated with \emph{no} adjustable parameters, equals 270\,nm, while the directly measured one is 320\,nm; the temperature dependence $\lambda(T)$ is also easily reproduced. Such agreement between the two completely different approaches allows us to conclude that ARPES studies of BKFA are bulk-representative. Our review of the available experimental studies of the superconducting gap in the new iron-based superconductors in general allows us to state that all hole-doped of them bear two nearly isotropic gaps with coupling constants $2\Delta/k_{\rm B}T_{\rm c}=2.5\pm1.5$ and $7\pm2$.
\end{abstract}


\maketitle
\section{Introduction}

The superconducting energy gap in the newly discovered iron-based superconductors naturally attracted much attention of physicists, and during one year of hard work, these materials were investigated by numerous experimental techniques \cite{Volodya, Volodya2, Evtushinsky, Chen, Wang, Szabo, Gonnelli, Gonnelli2, Samuely, Samuely2, Kaminski1, Ding, Zhou, Kaminski, Hasan, Terashima, Kawasaki, Ren, Hiraishi, Khasanov, Khasanov2, Hashimoto, Li, Dubroka, Millo, Boyer, Massee, Yin, Hinkov, Uemura}. As the diversity of conclusions made about the symmetry and value of the gap is huge, which can be attributed to the various shortcomings of different methods, the situation seems to be far from clear. In this paper, based on the angle resolved photoemission spectroscopy (ARPES) and muon spin rotation ($\mu$SR) data taken from the same single crystals of Ba$_{1-x}$K$_{x}$Fe$_2$As$_2$ with $T_\text{c}=32$\,K, we succeeded to reveal robust momentum dependence of the gap in this compound. The coupling constant, $2\Delta/k_{\rm B}T_{\rm c}$, is $\simeq1$ for the outer $\Gamma$-barrel and $6.8$ for all other Fermi surface sheets. Furthermore, close inspection of many studies of different iron-based superconductors allows one to derive quite definitive conclusions about the gap in these materials.

\section{The London penetration depth from the electronic band structure}
The London penetration depth, $\lambda$, can be expressed through the electronic band structure. For the quasi-two-dimensional superconductor with equivalent $a$ and $b$ principal axes, the formula, relating in-plane penetration depth to the band dispersion, reads (in SI units)
\begin{equation}
\frac{1}{\lambda^2(T)}=\frac{e^2}{2\pi\varepsilon_0c^2hL_c}\cdot\int\limits_{\text{FS}} v_\text{F}(\mathbf{k})\left[ 1 - \int\limits_{-\infty}^{+\infty} \left(-\frac{\partial f_T(\omega)}{\partial \omega} \right) \left|\text{Re}\frac{\omega+i\Sigma^{\prime\prime}}{\sqrt{(\omega+i\Sigma^{\prime\prime})^2-\Delta_{\mathbf{k}}^2(T)}}\right|{\rm d}\omega\right] {\rm d}k,
\end{equation}


where $v_\text{F}$ is the Fermi velocity, $\Delta_\mathbf{k}(T)$ is the momentum-dependent superconducting gap, $\Sigma^{\prime\prime}$ is the scattering rate (in the following we assume clean limit, $\Sigma^{\prime\prime}=0$), ${\rm d}k$ is the element of the Fermi surface length, $T$ is temperature, $L_c$ is the size of the elementary cell along the $c$ axis, $f_T(\omega)=[1+\operatorname{exp}(\omega/k_\text{B}T)]^{-1}$ is the Fermi function, $k_\text{B}$ is the Boltzmann constant, $h$ is the Planck's constant, $\varepsilon_0$ is the electric constant, $c$ is the speed of light, and $e$ is the elementary charge \cite{parabolic}. Formula (1) is consistent with results already presented in the literature \cite{Chandrasekhar}, although the former accounts for a finite lifetime (see also Ref.~\onlinecite{Evtushinsky}), and for the four-fold symmetry of the problem (see also Ref.~\onlinecite{Evtushinsky_Hall}).

\section{Low-energy electronic band structure of $\text{Ba}_{1-x}\text{K}_{x}\text{Fe}_2\text{As}_2$}
The information, required to calculate $\lambda(T)$  via the formula (1), can be extracted directly from ARPES spectra. The temperature and momentum dependence of the superconducting gap were obtained in Ref.~\onlinecite{Evtushinsky}, and the band structure was qualitatively revealed in Refs.~\onlinecite{Volodya} and \onlinecite{Volodya2}. The momentum dependence of the superconducting gap is quite easy to describe\,---\,the gap is large, $\Delta_{\mathbf{k}}(T)=\Delta_{\text{large}}(T)$, on the inner $\Gamma$-barrel and the propeller-like structure around the X point, and it is small, $\Delta_{\mathbf{k}}(T)=\Delta_{\text{small}}(T)$, on the outer $\Gamma$-barrel. The temperature dependence of the gap (see Fig.~1) is well fitted by the formula \cite{Gross_formula}
\begin{equation}
\Delta_{\text{large,small}}(T)=\Delta_{\text{large,small}}(0)\cdot \tanh \left( \frac{\pi}{2}\cdot\sqrt{\frac{T_\text{c}}{T}-1}\right)
\end{equation}
with $\Delta_{\text{large}}(0)=9.1$\,meV and $\Delta_{\text{small}}(0)<4$\,meV.

\begin{figure}[b]
\vspace{-0ex}\includegraphics[width=0.3\columnwidth]{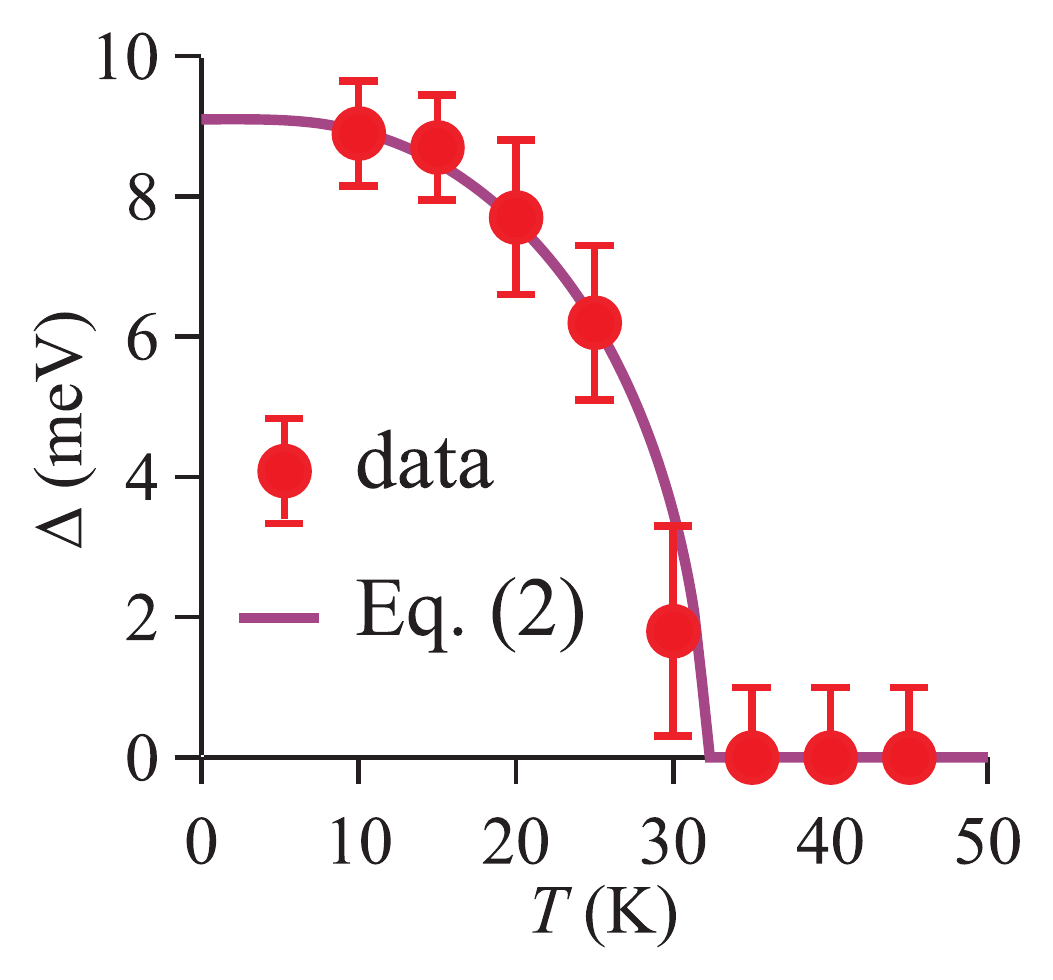}\vspace{-0ex}\caption{Temperature dependence of the superconducting gap in Ba$_{1-x}$K$_{x}$Fe$_2$As$_2$ as extracted from ARPES spectra \cite{Evtushinsky}. Underlying fitting curve is described by Eq.~(2).}
\vspace{-0ex}
\label{f:Model1}
\end{figure}

Taking into account the mentioned momentum dependence of the gap, one can rewrite (1) in the following way:

\begin{equation}
\frac{1}{\lambda^2(T)} = I_1\left[ 1 - D(\Delta_{\text{large}}(T), \Sigma^{\prime\prime}, T) \right] + I_2\left[ 1 - D(\Delta_{\text{small}}(T), \Sigma^{\prime\prime}, T)\right],
\end{equation}
where $I_{1,2}$ are temperature-independent factors
\begin{equation}
I_1 = \frac{e^2}{2\pi\varepsilon_0c^2hL_c} \int\limits_{\begin{smallmatrix} \text{outer }\Gamma, \\ \text{blades,} \\ \text{X-pocket} \end{smallmatrix}}\!\!\!\!\!\!\! v_\text{F}(\mathbf{k}) {\rm d}k, \quad \quad I_2 = \frac{e^2}{2\pi\varepsilon_0c^2hL_c} \int\limits_{\text{inner }\Gamma}\!\!\!\! v_\text{F}(\mathbf{k}) {\rm d}k,
\end{equation}
and $D(\Delta, \Sigma^{\prime\prime}, T)$ is defined as
\begin{equation}
D(\Delta, \Sigma^{\prime\prime}, T) \equiv \int\limits_{-\infty}^{+\infty} \left(-\frac{\partial f_T(\omega)}{\partial \omega}\right) \left|\text{Re}\frac{\omega+i\Sigma^{\prime\prime}}{\sqrt{(\omega+i\Sigma^{\prime\prime})^2-\Delta^2}}\right|{\rm d}\omega.
\end{equation}
See the Appendix for the evaluation of this integral.

\begin{figure}[]
\vspace{-0ex}\includegraphics[width=1.0\columnwidth]{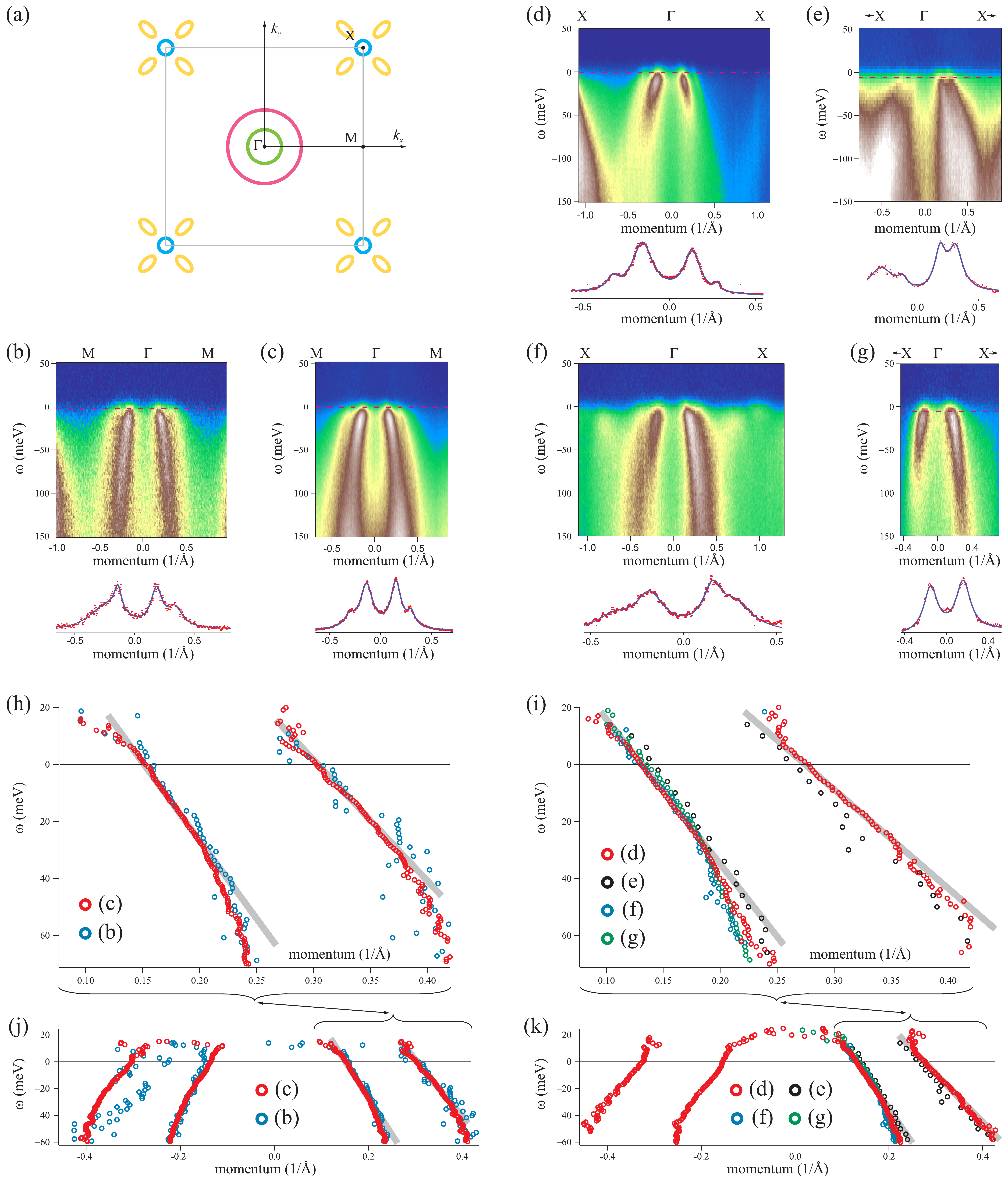}\vspace{-0ex}\caption{Determination of the low-energy band dispersion of Ba$_{1-x}$K$_{x}$Fe$_2$As$_2$ from ARPES spectra. (a) Fermi surface of Ba$_{1-x}$K$_{x}$Fe$_2$As$_2$, as seen in ARPES \cite{Volodya, Evtushinsky, Volodya2}. Cuts along $\Gamma$M: (b) $\Gamma_{-2,0}$ (index enumerates Mahan photoemission cones \cite{Mahan}), $T = 36$\,K, $h\nu=80$\,eV, horizontal polarization; (c) $\Gamma_{0,0}$, $T = 45$\,K, $h\nu=50$\,eV, horizontal polarization. Cuts along $\Gamma$X: (d) $\Gamma_{0,0}$, $T = 35$\,K, $h\nu=70$\,eV, vertical polarization; (e) $\Gamma_{+1,0}$, $T = 35$\,K, $h\nu=70$\,eV, vertical polarization; (f) $\Gamma_{0,0}$, $T = 35$\,K, $h\nu=80$\,eV, horizontal polarization; (g) $\Gamma_{0,0}$, $T = 41$\,K, $h\nu=40$\,eV, vertical polarization. Momentum distribution curves (MDC) taken nearly at the Fermi level are shown below each energy-momentum cut in order to demonstrate the high quality of the data and fits. Low-energy band dispersion of Ba$_{1-x}$K$_{x}$Fe$_2$As$_2$ was extracted by an MDC fit from the data taken at different experimental conditions. The Fermi velocities and Fermi momenta can be determined from panels (h--i) and (j--k) respectively.}
\vspace{-0ex}
\label{f:Model1}
\end{figure}

\begin{figure}[]
\vspace{-0ex}\includegraphics[width=0.4\columnwidth]{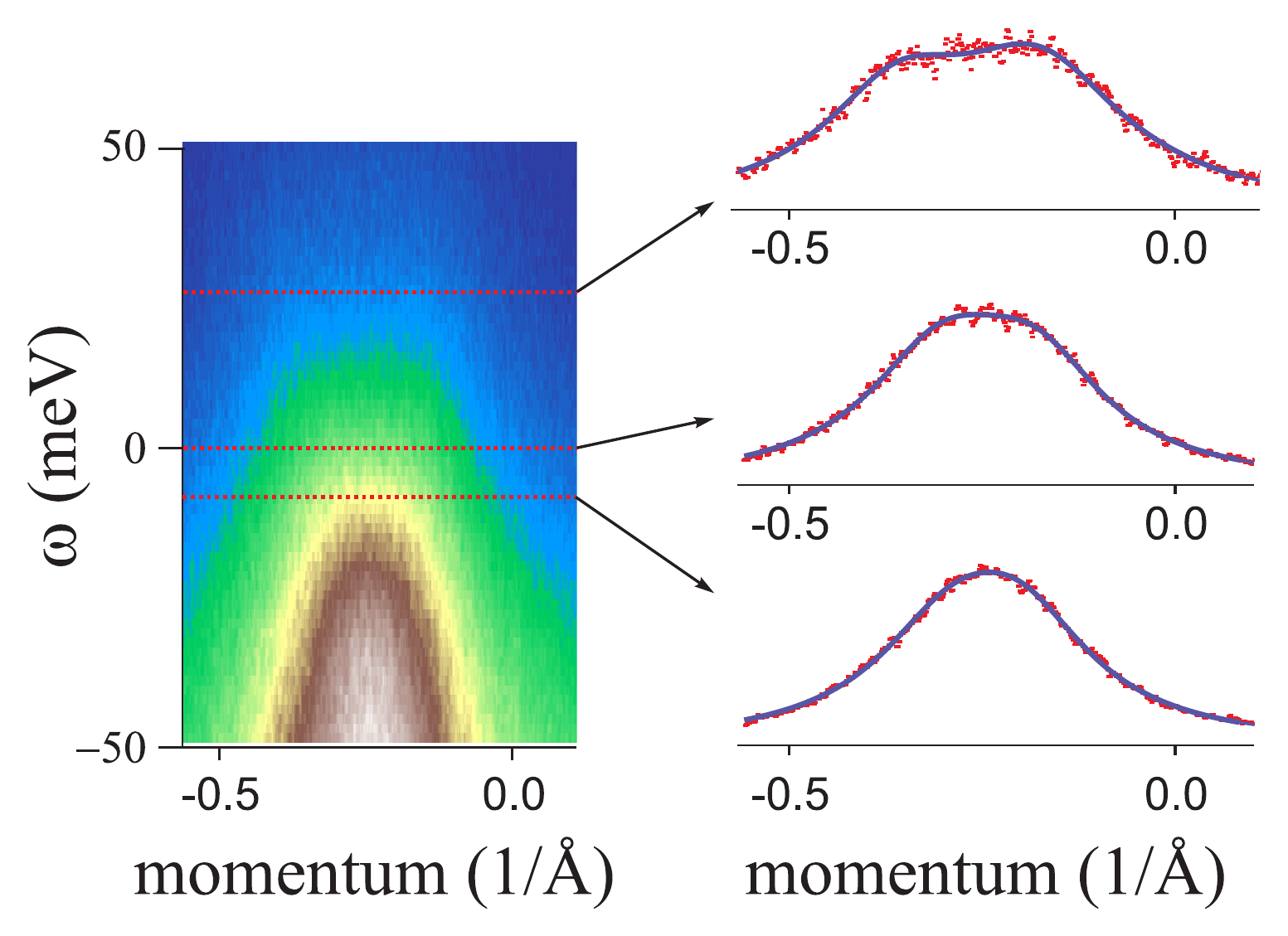}\vspace{-0ex}\caption{Dispersion of the X-pocket. The energy-momentum cut through the X point along $\Gamma$X, $T = 200$\,K (left on the figure) and fit of the MDC to two Lorentzians (right). Measurements at high temperatures allow us to track band dispersion of the X pocket also above the Fermi level. The experimental conditions are such that the blades are suppressed due to photoemission matrix elements effects.}
\vspace{-0ex}
\label{f:Model1}
\end{figure}

\begin{figure}[]
\vspace{-0ex}\includegraphics[width=0.85\columnwidth]{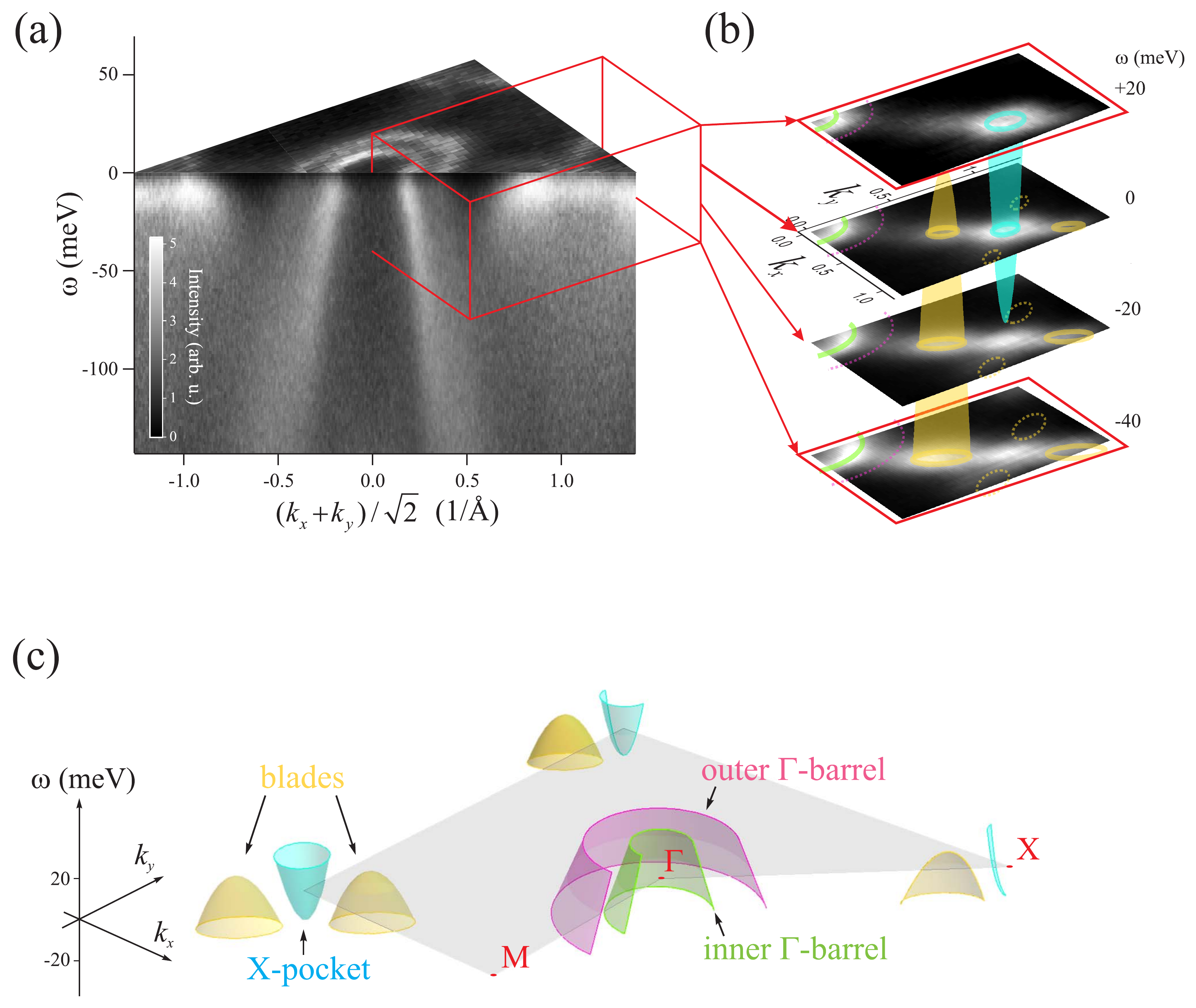}\vspace{-0ex}\caption{(a) A three-dimensional representation of the ARPES data. (b) The dispersion of the shallow bands near the X point can be determined from constant-energy cuts through the intensity distribution taken at different energies. The cross-section of the electron-like X pocket increases as with energy, while the cross-section of the hole-like blades decreases. (c) The model for the low-energy band dispersion in Ba$_{1-x}$K$_{x}$Fe$_2$As$_2$, derived from ARPES data [Fig.~2, Fig.~3, Fig.~4(b)].}
\vspace{-0ex}
\label{f:Model1}
\end{figure}

In Fig.~2 we present a quantitative investigation of the low-lying electronic band structure of Ba$_{1-x}$K$_{x}$Fe$_2$As$_2$. The band dispersion is extracted from ARPES data taken at a temperature slightly above the superconducting transition. The Fermi velocities for the inner and outer $\Gamma$-barrels along the $\Gamma$X direction equal $v^{\Gamma\text{X}}_{\text{i}\Gamma}=0.51$\,eV{\AA} and $v^{\Gamma\text{X}}_{\text{o}\Gamma}=0.36$\,eV{\AA} respectively [see Fig.~2(d--g), (i)], along the $\Gamma$M direction they are $v^{\Gamma\text{M}}_{\text{i}\Gamma}=0.54$\,eV{\AA} and $v^{\Gamma\text{M}}_{\text{o}\Gamma}=0.43$\,eV{\AA} [see Fig.~2(b), (c), (h)], resulting in $v_{\text{i}\Gamma}=0.52$\,eV{\AA} and $v_{\text{i}\Gamma}=0.40$\,eV{\AA} on the average. Fermi momenta for the inner and outer $\Gamma$-barrels are $k_{\text{i}\Gamma}=0.14$\,\AA$^{-1}$ and $k_{\text{o}\Gamma}=0.30$\,\AA$^{-1}$ respectively [see Fig.~2(b--g), (j--k)]. Kinks in the dispersions of the inner and outer $\Gamma$-barrels around 25\,meV have been described elsewhere \cite{Andreas}. For the electron-like X-pocket $k_\text{e}=0.06$\,\AA$^{-1}$, and the depth of the band is $\varepsilon_\text{e}=17\pm3$\,meV [see Fig.~3, Fig.~4(b)], from where, assuming a parabolic band dispersion for this small pocket, we infer $v_{\text{e}}=2\varepsilon_\text{e}/\hbar k_\text{e} = 0.57$\,eV{\AA}. For the hole-like blade pocket the average Fermi momentum equals $k_\text{h}=0.06$\,\AA$^{-1}$ and we estimate $\varepsilon_\text{h}$ as 5--15\,meV [see Fig.~4(b)], thus the average Fermi velocity equals $v_\text{h}=0.33$\,eV\AA. The $c$-axis lattice constant equals $13.3$\,\AA \cite{Rotter, Crystals}.

The low-energy band dispersion can be well approximated by the following formulas
\begin{enumerate}
    \item inner $\Gamma$-barrel
    \begin{equation}
        \xi_{\text{i}\Gamma}(k_x, k_y) =  0.52\left(0.14 - \sqrt{\left(k_x+\frac{2\pi m}{L_a}\right)^2 + \left(k_y + \frac{2\pi n}{L_a} \right)^2}\right), \quad m,n\in \mathbb{Z},
    \end{equation}
    where $L_a$ is the in-plane lattice constant, which according to Refs.~\onlinecite{Rotter, Crystals} equals $3.90$\,\AA;
    \item outer $\Gamma$-barrel
    \begin{equation}
        \xi_{\text{o}\Gamma}(k_x, k_y) =  0.40\left(0.3 - \sqrt{\left(k_x+\frac{2\pi m}{L_a}\right)^2 + \left(k_y + \frac{2\pi n}{L_a} \right)^2}\right), \quad m,n\in \mathbb{Z};
    \end{equation}

    \item X-pocket
    \begin{equation}
        \xi_{\text{X}}(k_x, k_y) =  0.017\left(\frac{\left[k_x+\frac{\pi(1+2m)}{L_a}\right]^2 + \left[k_y+\frac{\pi(1+2n)}{L_a}\right]^2}{0.06^2} - 1\right), \quad m,n\in \mathbb{Z};
    \end{equation}

    \item blades
    \begin{equation}
        \xi_{\text{b}}(k_x, k_y) =  0.01\left(1 - \left[\frac{\frac{k_x + k_y}{\sqrt{2}}+\frac{\sqrt{2}\pi(1+2m)}{L_a} \pm 0.36}{0.08}\right]^2 - \left[\frac{\frac{k_x - k_y}{\sqrt{2}}+\frac{\sqrt{2}\pi(1+2n)}{L_a}} {0.04}\right]^2\right), \quad m,n\in \mathbb{Z}.
    \end{equation}
\end{enumerate}
These dispersion relations are visualized in Fig.~4(c).

\section{Results}

The penetration depth at $T\rightarrow0$ in the clean limit depends only on the band structure and does not depend on the value of the superconducting gap (provided it is not zero), and, therefore, can be calculated purely from ARPES without any additional assumptions:

\begin{equation}
\frac{1}{\lambda^2(0)} = I_1 + I_2 = \frac{e^2}{2\pi\varepsilon_0c^2hL_c}\left[ ~~\int\limits_{\text{inner }\Gamma}\!\!\!\!\! v_\text{F}(\mathbf{k}) {\rm d}k + \int\limits_{\text{outer }\Gamma}\!\!\!\!\! v_\text{F}(\mathbf{k}) {\rm d}k + \int\limits_{\text{X-pocket }}\!\!\!\!\!\!\!\! v_\text{F}(\mathbf{k}) {\rm d}k + \int\limits_{\text{blades}}\!\!\!\!\! v_\text{F}(\mathbf{k}) {\rm d}k  \right],
\end{equation}
which results in $\lambda(0) = 270\text{\,nm}$.

This is in remarkable agreement with the value of 320\,nm obtained by $\mu$SR \cite{Khasanov}, the more so when one takes into account the complementarity of the two methods. The temperature dependence of $\lambda$ strongly depends on the values of the superconducting gap. Due to technical reasons, the small gap has not been determined precisely from ARPES measurements\,---\,only an upper limit of $\Delta_\text{small}<4$\,meV was obtained \cite{Evtushinsky}. Therefore, we use $\Delta_\text{small}$ as a fitting parameter when comparing $\lambda(T)$ calculated from ARPES to that determined from muon-spin depolarization rate in the $\mu$SR experiments (Fig.~5). The best fit of the normalized data corresponds to $\Delta_\text{small}=1.1$\,meV. The good agreement between absolute values of $\lambda$ at $T=0$ from ARPES (270\,nm) and $\mu$SR (320\,nm) implies correct determination of the band dispersion in the vicinity of the Fermi level. The possibility to fit the normalized temperature dependence with only one fitting parameter implies (i) correct determination of the relative contributions from different Fermi surface sheets, (ii) perfect agreement between two independent experimental techniques concerning the value of $\Delta_\text{large}$, and (iii) possibility to improve the estimate of $\Delta_\text{small}$ (now $2\Delta_\text{small}/k_{\rm B}T_{\rm c}\simeq1$) with respect to pure ARPES measurements ($<3$) \cite{Evtushinsky}. The general good agreement of ARPES and $\mu$SR studies of Ba$_{1-x}$K$_{x}$Fe$_2$As$_2$ allows us to state that ARPES experiments in this case are bulk-representative.

\begin{figure}[]
\vspace{-0ex}\includegraphics[width=0.5\columnwidth]{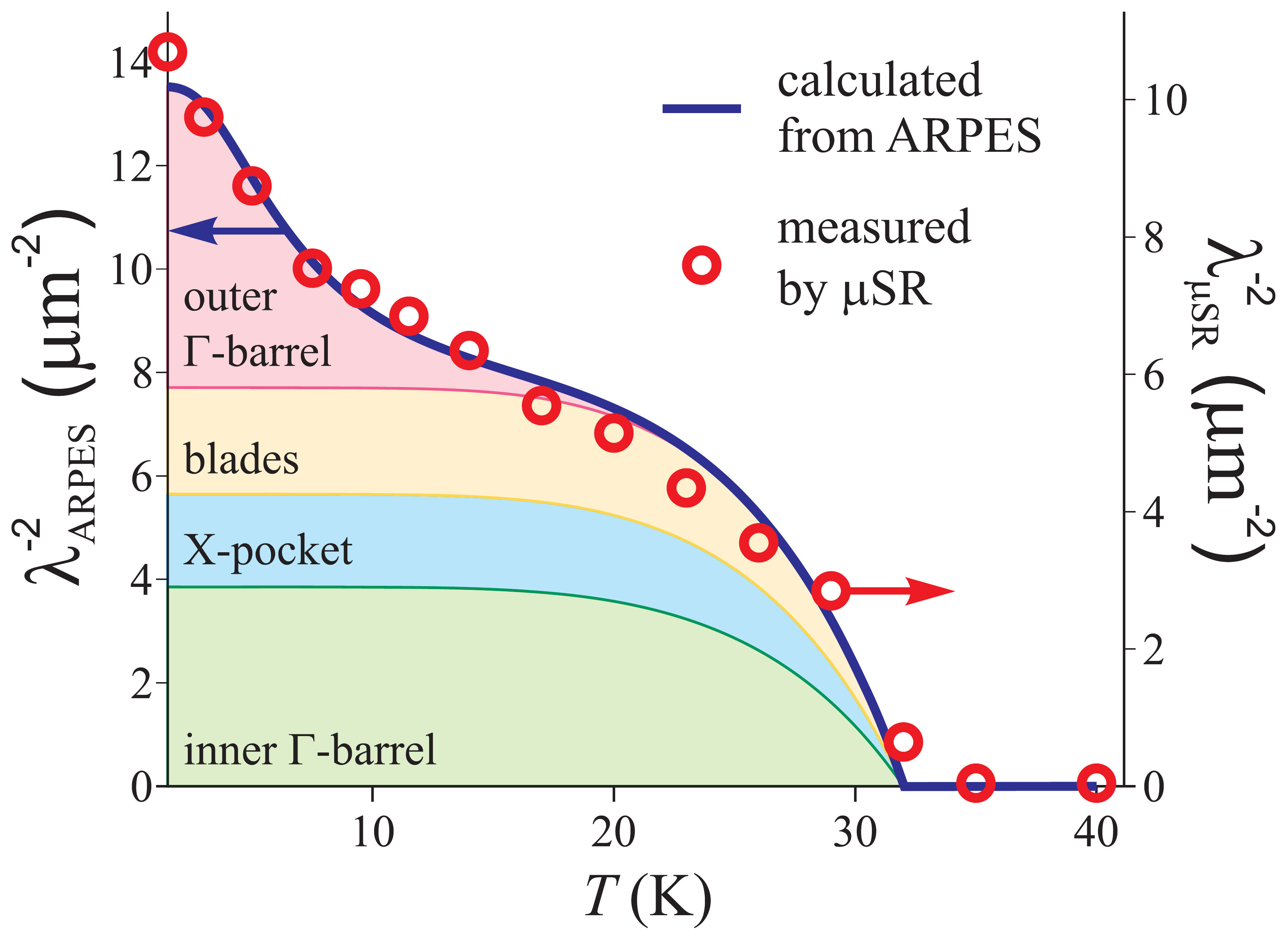}\vspace{-0ex}\caption{The in-plane London penetration depth in single crystals of Ba$_{1-x}$K$_{x}$Fe$_2$As$_2$ as calculated from ARPES with one adjustable parameter, $\Delta_{\text{small}}$, and as measured directly by $\mu$SR. The temperature dependence of the normalized penetration depth is reproduced with the best accuracy for $\Delta_{\text{small}}=1.1$\,meV, which is in agreement with our previous estimate $\Delta_{\text{small}}<4$\,meV \cite{Evtushinsky}. Contributions from different Fermi surface sheets are shown by different colors.}
\vspace{-0ex}
\label{f:Model1}
\end{figure}

\section{Experimental details}
Single crystals of Ba$_{1-x}$K$_{x}$Fe$_2$As$_2$ were grown using Sn as flux in a zirconia crucible. The growth details are described in Ref. \onlinecite{Crystals}. The crystals were cleaved \textit{in situ} and measured with Scienta SES R4000 analyzer at the base pressure of $5\cdot10^{-11}$\,mBar. ARPES experiments were performed using the ``$1^3$ ARPES'' end station at BESSY. Details of the experimental geometry can be found in Ref.~\onlinecite{Dima_geometry}.

$\mu$SR experiments were performed at the Swiss Muon Source (S$\mu$S),  Paul
Scherrer Institute (PSI, Switzerland).

\section{Overview of experimental studies of the gap in iron-based high-T$\text{c}$ superconductors}

\begin{figure}[!h]
\vspace{-0ex}\includegraphics[width=1\columnwidth]{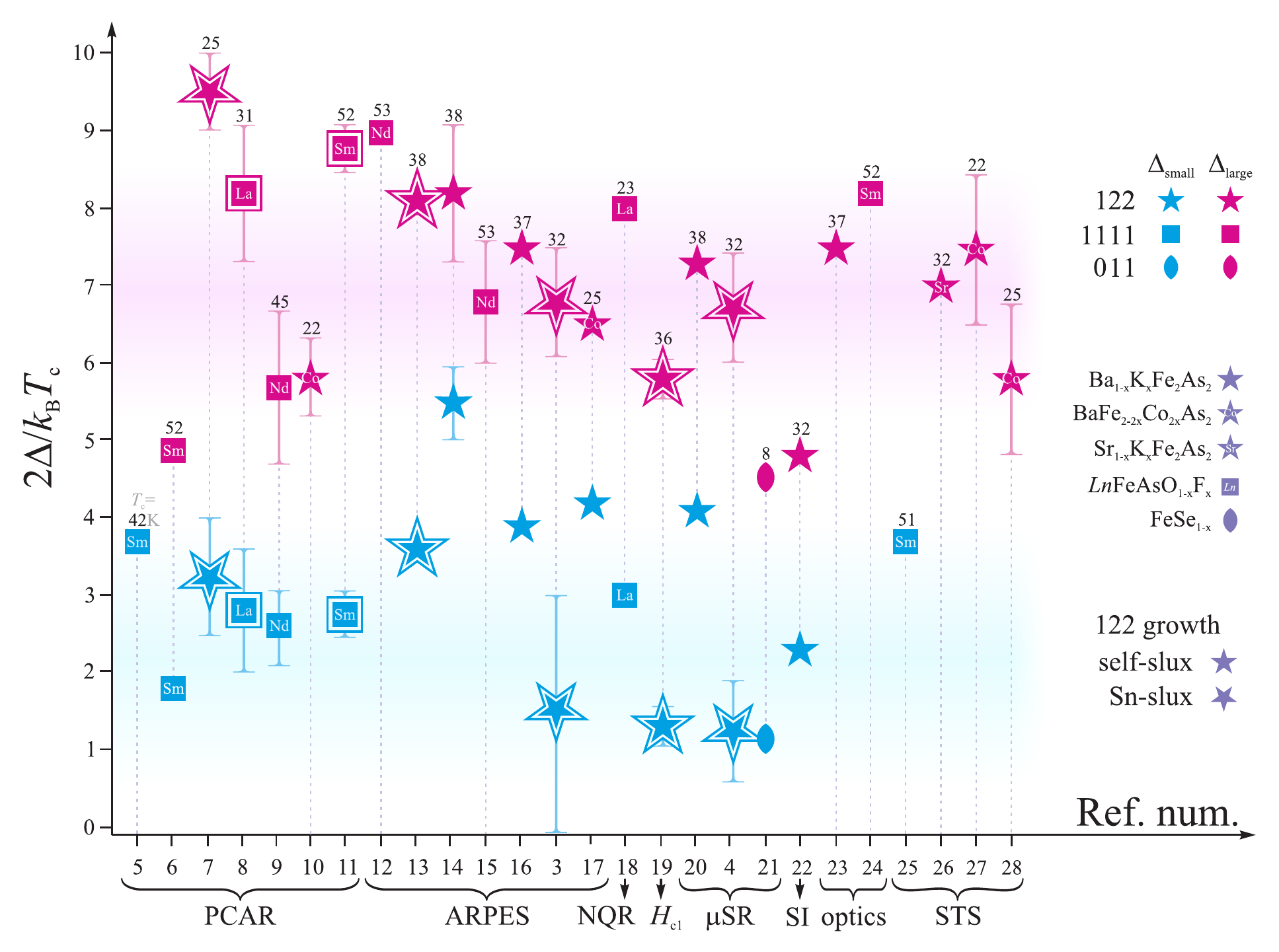}\vspace{-0ex}\caption{Coupling constant, $2\Delta/k_{\rm B}T_{\rm c}$, in iron arsenides, as revealed by different experimental techniques (refer to the main text for expansion of the abbreviations). In the figure, the points corresponding to the data taken on 122 systems are denoted by stars, points corresponding to 1111 systems are denoted by squares, and points corresponding to 011 (FeSe$_{0.85}$) are denoted by spindle-like symbols. Stars, corresponding to the studies of Sr$_{1-x}$K$_{x}$Fe$_2$As$_2$, are marked by ``Sr'', corresponding to BaFe$_{2-2x}$Co$_{2x}$As$_2$ are marked by ``Co''. For 1111 systems, the element $Ln$ in the structural formula $Ln$FeAsO$_{1-x}$F$_x$ is given inside the squares. Critical temperature, $T_{\rm c}$\,(K), is given as numbers above the symbols.  Blue symbols correspond to the small gap, while maroon ones correspond to the large gap. Studies on the 122 crystals grown by Sn-flux method are shown as overturned stars. Points corresponding to the most comprehensive and quality studies are marked by an extra frame. There are two superconducting gaps in these systems\,---\,the ``small'' one and the ``large'' one, although some studies overlook one of the gaps \cite{Kaminski1, Kaminski, Li, Dubroka, Boyer, Chen, Millo}.}
\vspace{-0ex}
\label{f:Model1}
\end{figure}

Extensive experimental studies, involving point contact Andreev reflection spectroscopy (PCAR) \cite{Chen, Wang, Szabo, Gonnelli, Samuely}, angle-resolved photoemission spectroscopy (ARPES) \cite{Kaminski1, Ding, Zhou, Kaminski, Hasan, Evtushinsky, Terashima}, nuclear quadrupole resonance (NQR) \cite{Kawasaki}, critical magnetic field ($H_{\rm c1}$) \cite{Ren}, muon spin rotation ($\mu$SR) \cite{Hiraishi, Khasanov}, surface impedance (SI) \cite{Hashimoto}, infrared spectroscopy (optics) \cite{Li, Dubroka}, and scanning tunneling spectroscopy (STS) \cite{Boyer, Millo} measurements carried out on different iron-based superconductors, SmFeAsO$_{1-x}$F$_x$ \cite{Chen, Gonnelli2, Dubroka, Wang, Millo}, LaFeAsO$_{1-x}$F$_x$ \cite{Gonnelli, Kawasaki}, NdFeAsO$_{1-x}$F$_x$ \cite{Samuely, Kaminski1, Kaminski}, Ba$_{1-x}$K$_{x}$Fe$_2$As$_2$ \cite{Szabo, Ding, Zhou, Hasan, Evtushinsky, Ren, Hiraishi, Khasanov, Hashimoto, Li}, BaFe$_{2-2x}$Co$_{2x}$As$_2$ \cite{Samuely2, Terashima, Massee, Yin}, Sr$_{1-x}$K$_{x}$Fe$_2$As$_2$ \cite{Boyer}, FeSe$_{0.85}$ \cite{Khasanov2}, let us conclude that these systems exhibit two superconducting gaps\,---\,a small one with the coupling constant $2\Delta_\text{small}/k_\text{B}T_\text{c}\sim2.5$, and a large one with $2\Delta_\text{large}/k_\text{B}T_\text{c}\sim7$ (see Table~I and Fig.~6). Some studies overlook one of the gaps (either small \cite{Kaminski1, Kaminski, Li, Dubroka, Boyer} or even the large one \cite{Chen, Millo}).

ARPES measurements allow one not only to state that different bands bear different gaps, but also to reveal the complete momentum dependence of the gap magnitude\,---\,in Ba$_{1-x}$K$_{x}$Fe$_2$As$_2$ the large gap opens on the inner $\Gamma$-barrel and the propeller-like structure around the X point, while the small gap opens only on the outer $\Gamma$-barrel \cite{Evtushinsky, Ding}. It is interesting to note that recent ARPES studies of the electron-doped compound BaFe$_{1.85}$Co$_{0.15}$As$_2$ have suggested that the smaller gap opens on the bands in the vicinity of X point, while the large one opens on the bands around $\Gamma$ \cite{Terashima}.
The anisotropy of the gap within one Fermi surface sheet has not been firmly established, although some evidence for small variations within the inner $\Gamma$-barrel ($\sim$10\%) was reported \cite{Zhou, Evtushinsky}. In addition, it is worthwhile noting that all of the above referred only to the magnitude (absolute value) of the gap. As suggested by NMR studies, the order parameter changes sign between different Fermi surface sheets \cite{Hajo}.

\begin{table}[]
\begin{tabular}{l@{~~~\,}|l@{~~}l@{~~}l@{~~}l@{~~}l@{~~}l@{~~}l@{~~}|l@{~~}l@{~~}l@{~~}l@{~~}l@{~~}l@{~~}l@{~~}|l@{~~}|l@{~~}|l@{~~}l@{~~}l@{~~}|l@{~~}|l@{~~}l@{~~}|l@{~~}l@{~~}l@{~~}l@{~~}}
\toprule
Method &\multicolumn{7}{c|}{PCAR~~} &\multicolumn{7}{c|}{ARPES~~} &\multicolumn{1}{l|}{NQR} &\multicolumn{1}{c|}{$H_{\rm c1}$} &\multicolumn{3}{c|}{$\mu$SR} &\multicolumn{1}{c|}{SI~} &\multicolumn{2}{c|}{optics~~} &\multicolumn{4}{c}{STS~~}\\
\hline
Ref. num. &\onlinecite{Chen} &\onlinecite{Wang} &\onlinecite{Szabo} &\onlinecite{Gonnelli} &\onlinecite{Samuely}  &\onlinecite{Samuely2} &\onlinecite{Gonnelli2} &\onlinecite{Kaminski1} &\onlinecite{Ding} &\onlinecite{Zhou} &\onlinecite{Kaminski} &\onlinecite{Hasan} &\onlinecite{Evtushinsky} & \onlinecite{Terashima} &\onlinecite{Kawasaki} &\onlinecite{Ren} &\onlinecite{Hiraishi} &\onlinecite{Khasanov} &\onlinecite{Khasanov2}& \onlinecite{Hashimoto} &\onlinecite{Li} &\onlinecite{Dubroka} &\onlinecite{Millo} &\onlinecite{Boyer} &\onlinecite{Massee} &\onlinecite{Yin}\\
\hline
Large gap &---&4.8&9.6&8.2&5.7&5.8&2.7& ~9 &8.1&8.2&6.8&7.5&6.8&6.5& ~8 &5.8&7.3&6.7&4.5&4.8&7.5&8.2&---& ~7&7.4&5.7 \\
Small gap &3.7&1.7&3.4&2.8&2.6&---&8.7&--- &3.6&5.5&---&3.9& <3&4.2& ~3 &1.3&4.1&1.2&1.1&2.3&---&---&3.7&---&---&---\\
\bottomrule
\end{tabular}
\caption{Coupling strength, $2\Delta/k_\text{B}T_\text{c}$, in iron-arsenic superconductors, as revealed by different experimental techniques\,---\,compare to the BCS universal value 3.53. Most of the available studies reveal two superconducting gaps of different magnitudes, which are represented in the table as ``large'' and ``small''. Refs.~\onlinecite{Chen, Wang, Szabo, Gonnelli, Samuely, Samuely2, Gonnelli2} are PCAR studies, Refs.~\onlinecite{Kaminski1, Ding, Zhou, Kaminski, Hasan, Evtushinsky, Terashima} are ARPES studies,  Ref.~\onlinecite{Kawasaki} is an NQR study, Ref.~\onlinecite{Ren} are critical magnetic field ($H_{\rm c1}$) measurements,  Refs.~\onlinecite{Hiraishi, Khasanov} are $\mu$SR studies, Ref.~\onlinecite{Hashimoto} is a surface impedance (SI) measurements, Refs.~\onlinecite{Li, Dubroka} are optics spectroscopy studies, and Refs.~\onlinecite{Millo, Boyer} are STS studies.}
\label{tab1}
\end{table}

\section{Conclusions}
In conclusion, we have derived low-energy electronic band structure  of Ba$_{1-x}$K$_{x}$Fe$_2$As$_2$ from ARPES spectra. Recently it was shown that ARPES allows one to explain and predict many tangible physical properties of the material, which depend on the low-lying electronic structure\,---\,transport properties \cite{Voigt, Evtushinsky_Hall, Narduzzo_Hall}, propensity of the system to form additional order \cite{TaSe2}, critical temperature of the superconducting transition \cite{Dahm}. In this paper we have presented a calculation of the London penetration depth from ARPES data (to the best of our knowledge, it is the first calculation of such kind). A comparison of the obtained results to direct $\mu$SR measurements has shown good agreement, which allows us to state that we have determined the robust momentum dependence of the superconducting gap in the bulk of Ba$_{1-x}$K$_{x}$Fe$_2$As$_2$. Namely, the gap distribution over the Fermi surface is consistent with those reported in our ARPES studies of this compound \cite{Evtushinsky}\,---\,the gap is small ($2\Delta_\text{small}/k_{\rm B}T_{\rm c} < 3$) on the outer $\Gamma$-barrel, and large on the other parts of the Fermi surface ($2\Delta_\text{large}/k_{\rm B}T_{\rm c} = 6.8$). Furthermore, comparison to $\mu$SR measurements resulted in the improvement of the assessment of the small gap magnitude\,---\,its coupling constant turned out to be $\simeq1$ instead of previous $<3$.
\section{Acknowledgements}
The project is part of the FOR538 and was supported by the DFG under Grants No. KN393/4, KN393/12, and BO1912/2-1. We thank A.\,N.~Yaresko for useful discussions, as well as R.\,Hübel, R.\,Schönfelder and S.\,Leger for technical support.

\section{Appendix}
Integrals of the form
\begin{equation}
D(\Delta, \Sigma^{\prime\prime}, T) = \int\limits_{-\infty}^{+\infty} \left(-\frac{\partial f_T(\omega)}{\partial \omega} \right) \left|\text{Re}\frac{\omega+i\Sigma^{\prime\prime}}{\sqrt{(\omega+i\Sigma^{\prime\prime})^2 -\Delta^2}}\right|\operatorname{d}\omega
\end{equation}
often appear upon calculation of the different physical properties of the materials from their low energy electronic structure. Unfortunately, the integration can not be performed analytically, therefore it is useful to find a convenient approximating formula. For the practically important case of $\Sigma^{\prime\prime}=0$, the function $D(\Delta, 0, T)$ can be approximated by an elementary function
\begin{equation}
M\left(\frac{\Delta}{k_\text{B}T}\right) = \frac{4}{\left(e^\frac{\Delta}{2k_\text{B}T} + e^{-\frac{\Delta}{2k_\text{B}T}} \right)^2}\sqrt{\frac{\pi}{8}\frac{\Delta}{k_\text{B}T} + \frac{1}{1 + \frac{\pi}{8}\frac{\Delta}{k_\text{B}T} } }.
\end{equation}

The accuracy of such approximation is better than 3\% for the entire range of parameters $\Delta$ and $T$:

\begin{equation}
-0.03<\frac{D(\Delta, 0, T) - M\left(\frac{\Delta}{k_\text{B}T}\right)}{D(\Delta, 0, T)} < 0.015 \quad\quad \forall \Delta, T > 0.
\end{equation}



\begin{thebibliography}{00}
\bibitem{parabolic} Note that in the simplest case of round Fermi surface and zero temperature, Eq.\,(1) yields a well known formula: $\lambda^2=\varepsilon_0mc^2/ne^2$ (or $\lambda^2=c^2m/4\pi ne^2$ in CGS units), where $n=(2\pi^2 L_c)^{-1} \int f(\epsilon(\mathbf{k})){\rm d}^2\mathbf{k}  $ and $m=\hbar k_\text{F}/v_\text{F}$
\end{thebibliography}
\end{document}